# On-Grid Doa Estimation Method Using Orthogonal Matching Pursuit

Abhishek Aich[†] and P. Palanisamy[*]

*Abstract*— Direction of Arrival (DOA) estimation of multiple narrow-band coherent or partially coherent sources is a major challenge in array signal processing. Though many subspace-based algorithms are available in literature, none of them tackle the problem of resolving coherent sources directly, e.g. without modifying the sample data covariance matrix. Compressive Sensing (CS) based sparse recovery algorithms are being applied as a novel technique to this area. In this paper, we introduce Orthogonal Matching Pursuit (OMP) to the DOA estimation problem. We demonstrate how a DOA estimation problem can be modelled for sparse recovery problem and then exploited using OMP to obtain the set of DOAs. Moreover, this algorithm uses only one snapshot to obtain the results. The simulation results demonstrate the validity and advantages of using OMP algorithm over the existing subspace- based algorithms.

*Keywords*—Array signal processing, Direction of Arrival, Orthogonal Matching Pursuit, Compressive Sensing

## I. INTRODUCTION

Direction of Arrival estimation problem is one of the classic field of study in the area of array signal processing field. To solve this problem, researchers have proposed many classes of algorithms such as subspace-based algorithms like the multiple signal classification (MUSIC) [1] and the estimation of parameters by rotational invariant techniques (ESPRIT) [2], the nonlinear least squares (NLS) method, commonly known as the maximum likelihood estimation. Unfortunately, these algorithms come with certain limitations e.g. the subspace-based algorithms and the NLS need to have a priori knowledge of the source number. Additionally, subspace-based algorithms require to compute sample data covariance matrix and hence these require a sufficiently large number of snapshots. Another major drawback is when there is source coherency which lead to cause a rank deficiency in the covariance matrix. Though algorithms like maximum likelihood algorithm [3] and weighted subspace fitting algorithm [4] solve the case of coherent sources, their time complexity is high as they involve a multiple dimensional search and they have no guarantees of global convergence as they require good initial estimates.

Sparsity property of signal sources in *angle space* or sources in spatial domain make it convenient for us to utilize compressive sensing (CS) [8] [16] as a tool to solve DOA estimation problem. The sparse recovery algorithms of CS are now being applied in several scenarios recently such as cases with no knowledge of the source number, limited number of snapshots (even a single snapshot), and completely or partially coherent sources. Because of these interesting properties, the link between array signal processing and CS theory has gained interest for further studies [5] [6]. We would like to point out a very important key difference between the common sparse representation data model and DOA estimation data model used in array signal processing. The literature survey on sparse representation model reveals that it has always been confined to discrete linear systems. Contrary to this, the DOA attributes are continuous in nature and the received data is nonlinear in the DOAs. Therefore, the sparse model representation for DOA estimation can be classified into three categories based on the model adopted- on-grid, off-grid and gridless. For on-grid model representation, the data model is obtained with the assumption that the true DOAs lie on a set of fixed grid points in order to directly apply the existing sparse recovery algorithms. For off-grid model representation, the DOAs are not confined to be on the grid, though a grid is still required. Finally, the gridless model representation which is recently developed, do not need a grid and can work in the continuous domain directly.

CS sparse signal reconstruction algorithms can be broadly classified into two categories – Convex optimization based algorithms and the class of greedy algorithms. In this paper, we concentrate on a greedy algorithm called as orthogonal matching pursuit (OMP) [7]. The typical greedy sparse recovery approaches are basis pursuit (BP), compressive sampling matching pursuit (CoSaMP), orthogonal matching pursuit (OMP) and sparse Bayesian learning (SBL) [9]. Among these approaches, OMP is a compelling and favorable candidate for solving the DOA estimation problem due to its attractive properties. The main advantages of this algorithm are its low computational complexity and speed of recovery. Also, in addition to its fast implementation, OMP is also empirically advantageous in terms of recovery performance [10]. The subspace-based algorithms have considerably huge computational cost (estimation of covariance matrix and eigen decomposition), memory cost (large number of snapshots) and additionally, they require known or estimated noise level a prior which creates inconvenience for real time applications. Here's where OMP fits in the DOA estimation scenario. It has both low computational complexity as well as efficient applicability for redundant dictionary. It requires no knowledge of noise level and inherently solves the coherency problem. Thus, this makes OMP a suitable candidate for engineering practice.

The reminder of the paper is organized as follows. The system model and sparse representation of DOA estimation problem is presented in Section II. The application OMP algorithm to DOA estimation is provided in Section III. Section

The authors are with the Department of Electronics and Communication Engineering, National Institute of Technology, Tiruchirappalli, 620 015, India.

e-mail: [†]abhishekaich.nitt@gmail.com, [*]palan@nitt.edu

IV demonstrates the performance and effectiveness of the OMP algorithm and Section V concludes the paper.

## II. PRILIMINARIES

### A. Data Model

Consider $M$ narrowband far-field source signals $s_i$, $i = 1, 2, \ldots, M$ impinging on a uniform linear array (ULA) of $N$ omnidirectional sensor from directions $\theta_i$, $i = 1, 2, \ldots, M$. The time delays at different sensors are represented by simple phase shifts which results in the following model:

$$x(t) = \sum_{i=1}^{M} a(\theta_i)s_i(t) + n(t); \; t = 1, 2, \ldots, K$$

$$= A(\theta)s(t) + n(t); \; t = 1, 2, \ldots, K \quad (1)$$

here $t$ represents the snapshot $K$. Here, $x(t) \in \mathbb{C}^N$, $s(t) \in \mathbb{C}^M$ and $n(t) \in \mathbb{C}^N$ represent the received data, source signal vector and the noise vector at snapshot time $t$ respectively. $a(\theta_i)$ is the steering vector of the respective $i^{th}$ source which is also dependent on the geometry of the sensor array. These form the array manifold matrix $A(\theta)$ consisting of all the steering vectors $a(\theta_i)$, $i = 1, 2, \ldots, M$. In matrix form, (1) can be written as

$$X = A(\theta)S + N \quad (2)$$

where $X = [x(1), x(2), \ldots, x(K)]$, $S = [s(1), s(2), \ldots, s(K)]$ and $N = [n(1), n(1), \ldots, n(K)]$. The objective is to estimate the parameters $\theta_i$, $i = 1, 2, \ldots, M$ that are referred to as the DOAs given $X$ and the mapping $\theta \to a(\theta)$. In practice, the source number $M$ is usually unknown and is assumed to be smaller than $N$ for the DOAs to be uniquely identified. The condition for identifying these sources is explained in the following section.

### B. Parameter Identifiability in ULA

Let $\P_\theta$ denote the domain of the DOAs that can be scanned by the array $[0°, 180°]$. Define a set

$$\Lambda := \{ a(\theta) : \theta \in \P_\theta \}$$

Then for a $N$-element ULA, the following condition holds

$$spark(\Lambda) = N + 1 \quad (3)$$

where spark $(\Lambda)$ is defined as smallest number of elements in $\Lambda$ that are linearly dependent. In [11], it is shown that any $M$ sources can be uniquely identified from (2) provided that

$$M < \frac{spark(\Lambda) - 1 + rank(S)}{2} \quad (4)$$

But, since (4) requires knowledge of $S$, it can be simplified [12] to

$$M < \frac{spark(\Lambda) - 1 + rank(X)}{2} \quad (5)$$

In case of ULA, this further reduces to

$$M < \frac{N + rank(X)}{2} \quad (6)$$

(6) gives the condition required to guarantee unique identifiability for any $M$ source signals.

### C. Sparse Representation for Compressive Sensing

Now we present the sparse representation of (1) in order to utilize CS based recovery algorithms for obtaining the DOAs. Let $y \in \mathbb{C}^m$ be the measured vector obtained after multiplying $x \in \mathbb{C}^N$ (1 snapshot) with a measurement matrix $\phi \in \mathbb{C}^{m \times N}$. Then,

$$y = \phi x \quad (7)$$
$$\Rightarrow \quad y = \phi A(\theta)s + \phi n \quad (8)$$

$y$ forms the sparse representation of received data $x$. The underlying principle for sparse representation is that though the received data $x$ lies in a high-dimensional space, it can be well approximated in some lower-dimensional subspace ($m < N$) where m $\geq M\ln(N)$. In DOA estimation, the data snapshot $y$ is a linear combination of the steering vectors $a(\theta_i)$, of the sources $\theta_i$, $i = 1, 2, \ldots, M$ and the sparsity infers from the fact that the sources are less than the sensors ($M < N$). Hence, given $y$ and $\phi A$, the problem of sparse representation is to find the vector $s$ subject to data consistency. Intuitively, we should find the sparsest solution. This can be done by solving the following optimization problem (considering noiseless case):

$$\min \|s\|_0 \; s.t. \; y = \phi A s \quad (9)$$

where $\|s\|_0 := \# \{j : s_j \neq 0\}$ gives the total number of non-zero entries of $s$ referred as sparsity of $s$. In this paper, we do so by using the greedy algorithm OMP.

### D. Class of on-grid sparse recove algorithm

These algorithms are developed by directly applying compressed sensing techniques to DOA estimation. The DOAs are assumed to lie on a prescribed grid (hence called *on-grid*) so that the objective problem can be solved CS based sparse recovery algorithms. So, to connect the continuous DOA estimation and discrete CS representation, assume that for the on-grid sparse algorithms, the continuous DOA domain $\P_\theta$ can be replaced by a given set of grid points

$$\Omega := \{\Omega_1, \Omega_2, \ldots, \Omega_{N_s}\} \quad (10)$$

where $N_s \gg N$ denotes the domain size or grid size. Therefore, the required DOAs can only take values in $\Omega$. We can now express the $N \times N_s$ basis matrix $A$ for $s$ as

$$A(\Omega) := \{A(\Omega_1), A(\Omega_2), \ldots, A(\Omega_{N_s})\} \quad (11)$$

From above form of $A$, it can be observed that for each $t$, $x$ contains only $M$ nonzero values, whose locations correspond to the $M$ DOAs, and therefore it is said to be a sparse vector as $N_s \gg M$. Note that we are considering the case of a single snapshot i.e. $K = 1$. Thus, it is a single measurement vector (SMV) sparse representation and can be directly applied for DOA estimation. In case of multiple measurement vector (MMV) sparse representation, $K > 1$ and it will be a case where we need to exploit the temporal redundancy of these snapshots.

## III. ANALYSIS OF OMP ALGORITHM FOR DOA ESTIMATION PROBLEM

The OMP algorithm was originally proposed for general sparse recovery [7]. In this paper, we introduce the OMP

algorithm for DOA estimation scenario. Let us first explain the underlying computational technique of the algorithm and connect the CS based terminology with DOA estimation terminology. $s$ is the $M$-sparse signal of interest, sparse in basis $\mathbf{A}$ (12). The measurement matrix is referred as $\phi$ and measured vector is $y$. Since $s$ has only $M$ source information, the measured data vector (considering noiseless case)

$$y = \phi \mathbf{A}\, s \quad (12)$$
$$y = \Psi s \quad (13)$$

Here, $y$ is a linear combination of $M$ columns of $\Psi\ (= \phi \mathbf{A})$. In CS theory terminology, $y$ has an $M$-term representation over the dictionary $\Psi$. This analogy allows transportation of DOA estimation problem to the CS signal recovery problem. $c$ is the current iteration number and $\Lambda_c$ is the support set at $c^{th}$ iteration.

Now, we explain the idea behind OMP algorithm for DOA estimation. For this we need to determine $s$. To obtain $s$, we need to find which columns of $\Psi$, defined as $\gamma_i$ ; $i = 1, 2, \ldots, N_s$, are responsible for creating the measured vector $y$. This is done in a greedy fashion (hence the name *greedy* algorithm). At every iteration, choose the column of $\Psi$ that is most strongly correlated with the remaining part of $y$. Then we remove its contribution to $y$ by subtraction and iterate on the residual left. Then accordingly, after $M$ iterations, the algorithm should have identified the correct set of corresponding columns. The algorithm is as follows:

---
**Algorithm:** OMP for DOA estimation

**Input**:
- $\Psi$, $\mathbf{y}$, M

**Output**:
- An estimate $\hat{s}$

**Procedure**:
1) Set $r_0 = y$, $\Psi_0 = \emptyset$, $\Lambda_0 = \emptyset$, and an iteration counter $c = 1$
2) Find the corresponding index $\lambda_c$ of the optimization problem

$$\lambda_c = \arg\max_{j \in \{1,\ldots,N\}} |\langle r_{c-1}, \gamma_j \rangle|$$

3) Augment the index set $\Lambda_0 = \Lambda_{c-1} \cup \{\lambda_c\}$ and the matrix of chosen atoms $\Psi_c = [\Psi_c\ \gamma_j]$
4) Solve the following optimization problem to obtain the signal vector estimate for $\Psi_c$:

$$\mathbf{s_c} = \arg\min_{\mathbf{s}} \|\Psi_c \mathbf{s} - \mathbf{y}\|_2$$

5) Calculate the new approximation ($\beta_c$) of $y$ and the new residual:

$$\beta_c = \Psi_c\, s_c$$
$$r_c = y - \beta_c$$

6) Increase $c$ by 1, and return to Step 2) if $c < M$.
7) Value of estimate $\hat{s}$ in $\lambda_j$ equals the $j^{th}$ component of $s_c$.

---

Also note that residual $r_c$ is always orthogonal to the columns of $\Psi_c$. Therefore, OMP always selects a new column at each step and $\Psi_c$ has full column rank. Then the peaks from the angle spectrum $\mathbf{P}_s(\theta)$ of $\hat{s}$ correspond to respective DOA, where

$$\mathbf{P}_s(\theta) = \|\hat{s}_\theta\|^2 ; \theta = \theta_1, \theta_2, \ldots, \theta_{N_s} \quad (14)$$

where $N_s$ being the total number of angles to be scanned.

## IV. NUMERICAL ANALYSIS

The following **MATLAB** simulations are being presented to study the performance of OMP algorithm in various DOA environment scenarios. Results are discussed subsequently with respective plots and analysis. The ULA is chosen with $N = 15$ sensors having an inter-sensor spacing of half a wavelength. The scanning direction grid contains 181 points being sampled from $-90°$ to $90°$ having $1°$ interval. Throughout this section, the noises are generated from a zero-mean complex Gaussian distribution.

**Simulation I:** In this example, we consider three non-coherent sources with respective directions as $\theta_1 = -40°$, $\theta_2 = 0°$ and $\theta_3 = 24°$. The number of snapshots is set as $K = 500$ for the subspace-based algorithms [1], [13] and [14] and $K = 1$ for the OMP algorithm. The SNR is set to be 0 dB. The performances of the algorithms are shown in Fig.1. For all the algorithms, Fig. 1 displays the normalized spatial spectra. We can observe from the plot that the OMP algorithm correctly resolves all the sources under the given scenario with better resolution. Although the mentioned subspace-based algorithms succeed to generate three visible peaks, a deviation between the estimated and true DOAs can be observed. This is mainly caused by the highly noisy environment.

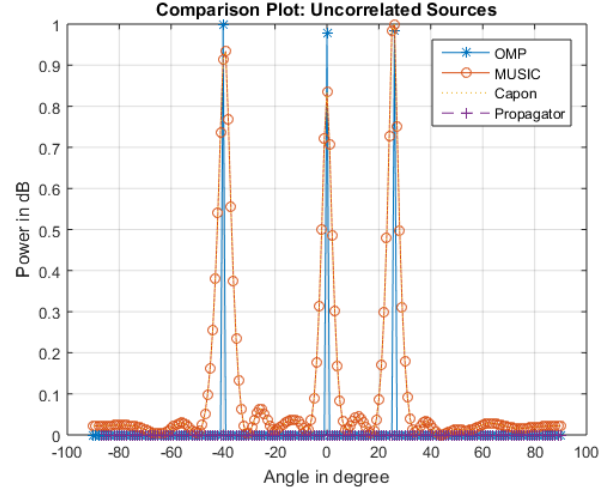

Figure 1. Plot for Simulation I (N = 15, M = 3, SNR= 0 dB)

**Simulation II:** Similar to the first example, we consider a source from $\theta_1 = -40°$ and two coherent sources with directions $\theta_2 = 1°$ and $\theta_3 = -24°$. The number of snapshots and SNR remain unchanged. The performances of the algorithms are compared in this partially coherent source environment. We again normalize the spectrum for proper comparison, shown in Fig.2. We can observe from the plot that the OMP algorithm accurately resolves all the sources under the partially coherent source scenario without false peaks. However, the subspace-based algorithms fail to resolve all the signals and also generate false peaks. This is because they suffer from rank deficiency of covariance matrix in case of coherent signals. For these subspace-based methods to work properly, some pre-processing of covariance matrix like spatial-smoothening need to be performed. Thus, OMP algorithm is independent of the coherency of the sources and doesn't require pre-processing of data for getting correct results. Again, a single snapshot is enough to resolve the DOAs unlike other algorithms which require multiple snapshots. This lowers the memory requirement of the data.

with 3 sources, SNR = 0 dB, and normalize and scale them with number of trial. It can be is observed from Fig. 4 that the OMP algorithm shows a major drawback in consistency and need to be improved for better performances in real time. Such a result can be attributed to the fact that the measurement matrix follows a strict restricted isometry property (RIP) for obtaining a correct result and hence, and hence the estimation is not independent of the selection of the columns of $\Psi$ [15].

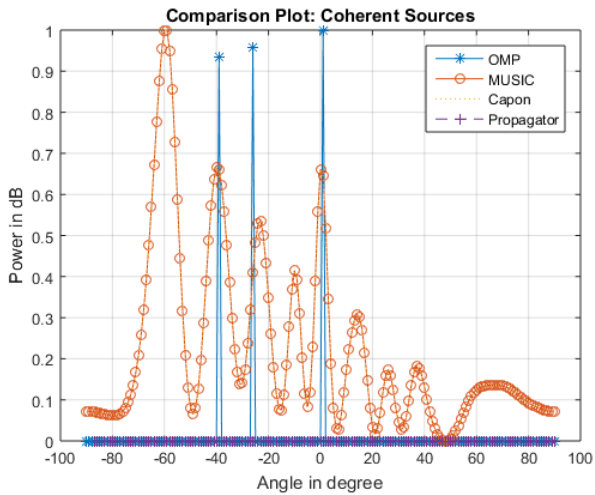

Figure 2. Plot for Simulation II (N = 15, M = 3, SNR= 0 dB)

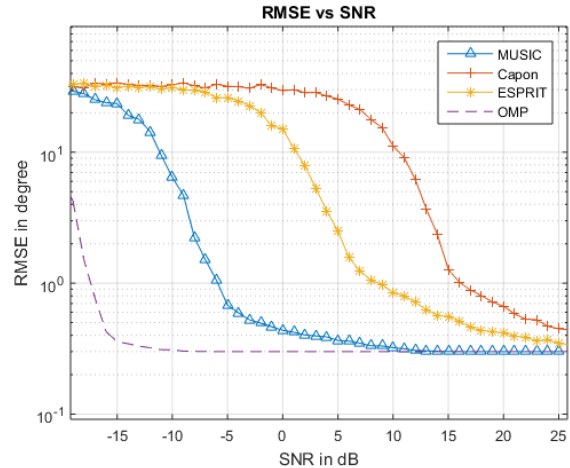

Figure 3. Plot for Simulation III (N = 15, M = 2, 1000 trials)

**Simulation III:** In this simulation, we consider two sources with respective directions as $\theta_1 = -50°$ and $\theta_2 = 60°$. We compare the OMP algorithm with [1], [2] and [13] in terms of root mean square error (RMSE) with respect to SNR (dB). The number of snapshots for them is $K = 1000$ for the subspace-based algorithms and use a single snapshot for the OMP algorithm. We perform 1000 independent Monte Carlo experiments for root mean square error (RMSE) for the DOAs estimation. It is observed from Fig. 3 that the OMP algorithm achieves a much better estimation performance at low SNRs. The subspace-based cannot work properly under the highly noisy environment. We observe that the OMP algorithm outperforms the other algorithms even in high SNR region.

## V. CONCLUSION

In this paper, we provided an overview of the application of a robust greedy compressive sensing algorithm, *i.e.* the OMP algorithm. The DOA estimation problem is modelled as a standard compressive sensing recovery problem. The most favorable advantage of the algorithm is that it does not require any eigen value decomposition and works with single snapshot. This is highly desirable for practical engineering applications such as dynamic tracking of a vehicle. On-grid effects on the DOA RMSE for the subspace-based algorithms and the OMP algorithm are compared which show the empirical advantage of OMP algorithm, when compared as a function of the SNR. Future research work will be to do an exhaustive performance analysis of the other greedy algorithms to DOA estimation scenario and study their performance bounds.

**Simulation IV:** In this simulation, we will test the consistency of the OMP algorithm and do 5 Monte Carlo trials

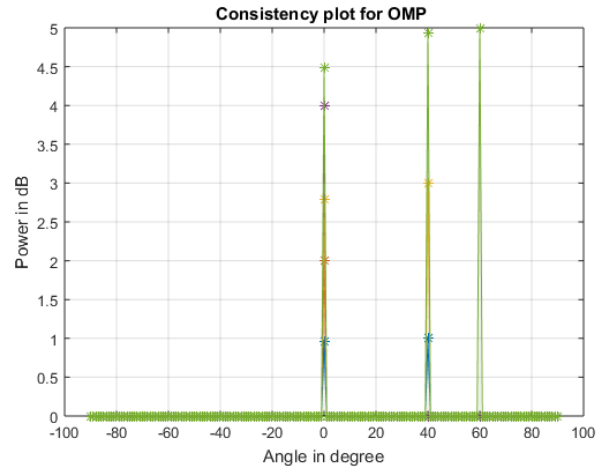

Figure 4. Plot for Simulation IV (N = 15, M = 3, 5 trials)